\documentclass[12pt,showpacs,preprintnumbers,amssymb]{revtex4}
\usepackage{epsfig}
\usepackage{bm}
\usepackage{amsmath}

\begin{document}

\title{  
First order phase transitions: 
equivalence between bimodalities and the Yang-Lee theorem }
\author{Ph.Chomaz$^{(1)}$,F.Gulminelli$^{(2)}$}
\affiliation
{(1) GANIL (DSM-CEA/IN2P3-CNRS), B.P.5027, F-14021 Caen C\'{e}dex, France\\
(2) LPC Caen (IN2P3-CNRS/ISMRA et Universit\'{e}), F-14050 Caen C\'{e}dex, France}
\begin{abstract}
First order phase transitions in finite systems can be 
defined through the  bimodality of the distribution 
of the order parameter. This definition is equivalent 
to the one based on the inverted curvature 
of the thermodynamic potential. Moreover we show that it is  
 in a one to one correspondence with the Yang Lee 
theorem in the thermodynamic limit. Bimodality is a 
necessary and sufficient condition for zeroes of the
 partition sum in the control intensive variable complex plane  
to be distributed on a line perpendicular 
to the real axis with a uniform density, 
scaling like the number of particles.
\end{abstract}

 
\maketitle

\smallskip Phase transitions are universal phenomena which have been
theoretically understood at the thermodynamic limit of infinite systems as
anomalies in the associated equation of state (EoS). They have been
classified according to the degree of non-analyticity of the thermodynamic
potential at the transition point. For finite systems, it is generally
believed that phase transitions cannot be defined since thermodynamic
potentials of a finite system in a finite volume are analytic functions.
This situation can be thought as unsatisfactory since on one hand the
thermodynamic limit does not exist in nature, and on the other hand many
experimental efforts are devoted to understand the link between mesoscopic
and macroscopic systems, especially for spectacular properties like phase
transitions~\cite{wales}.

A classification scheme valid for finite systems has been proposed by
Grossmann~\cite{grossmann} using the distribution of zeroes of the canonical
partition sum in the complex temperature plane. Alternatively, it has been
claimed that first order phase transitions in finite systems can be related
to a negative microcanonical heat capacity~\cite{gauss,gross} or more
generally to an inverted curvature of the thermodynamic potential as a
function of an observable which can then be seen as an order parameter~\cite
{prl99}. These anomalies can also be connected to the general topology of
the potential energy surface~\cite{pettini}. These very different approaches
share the common viewpoint that at the mesoscopic or microscopic scale phase
transitions are not smeared to become loosely defined smooth state changes,
but can be analyzed in a rigorous way without explicitly addressing the
thermodynamic limit. If this is certainly important to progress in the
experimental study of phase transitions in finite systems, it is also true
that the connection between these mesoscopic phenomena and the
thermodynamics of the macroscopic world is not clear. This is especially
annoying since the definitions of phase transitions in finite systems are
not in general equivalent and different conclusions can be drawn out of the
observation of a physical phenomenon depending on the theoretical framework
used. In particular it can be shown~\cite{deuteron} that any generic channel
opening may lead to a convexity anomaly of the microcanonical entropy which
does not necessarily survive to the bulk. In the same way the distribution
of zeroes close to the real axis may suggest a first order phase transition
even for systems where the transition temperature diverges in the
thermodynamic limit~\cite{isomers}. In order to have a coherent picture of
the physical meaning of a phase transition in a finite system it is then
 important to analyze in detail the conditions of persistency of the
signals towards the bulk and to make a clear bridge between the various
definitions.

We have recently proposed the possible bimodality of the probability
distribution of observable quantities as a connection between these 
ideas~\cite{topology}. In this letter we further demonstrate 
that this definition
is consistent with the Yang Lee theorem~\cite{yanglee}, $i.e.$ with the
standard definition of first order phase transitions in the thermodynamic
limit. Since our definition is a generalization of the previous ones~\cite
{grossmann,gauss,gross}, 
the conditions of compatibility of the different signals
and criteria proposed to identify first order phase transitions in finite
systems can be explicitly worked out.

Let us consider that for each event $i$ we can measure a set of $K$
independent observables, $b_{k}^{(i)},$ which form a space containing one
possible order parameter. We can sort events according to 
the results of  the measurements 
$ \vec {b}^{(i)}{\equiv }\{b_{k}^{(i)}\}$  and thus define
a probability distribution $P (\vec {b}) $. In the absence of a
phase transition $\log P (\vec {b}) $ is expected to be
concave. An abnormal (e.g. bimodal) behavior of $P (  \vec {b} ) $
or a convexity anomaly signals a first order phase transition~\cite{topology}%
.

This definition does not imply an a priori knowledge of the thermodynamic
potential, $i.e.$ of the properties of the specific equilibrium reached in
the experimental situation, but only requires the measurement of the event
distribution. Moreover, it does not assume the a-priori knowledge of an
order parameter; the order parameter can be extracted from the data
themselves by looking at the best direction in the observable space to split
the event cloud into two components. This topological definition is
therefore easy to experimentally implement, and in fact it has already
allowed to successfully recognize phase transitions in metallic clusters~\cite
{haberland} and in the multifragmentation of atomic nuclei~\cite{lopezo}.

Moreover this definition has a wider application domain than the previous
ones~\cite{grossmann,gauss,gross}; it can be used out of canonical or
microcanonical equilibria, for open systems~\cite{topology} and
non-extensive statistics such as Tsallis equilibria~\cite{next2001}. In
particular it can be viewed as an extension of the definition based on the
convexity anomaly of the microcanonical entropy~\cite{gauss,gross} or more
generally of any thermodynamical Gibbs potential~\cite{prl99}. Let us first
discuss Boltzmann-Gibbs equilibrium which is obtained by maximizing the
Shannon information entropy under the constraints of the various observables 
$b_{k}$ known in average. Then $P (  \vec {b} ) $ can be written
as 
\begin{equation}
\log P_{{ \vec {\lambda} }} (  \vec {b} ) =\log \bar{W} ( 
 \vec {b} ) -\sum_{k=1}^{K}\lambda _{k}b_{k}-\log Z_{ \vec {\lambda }}
\label{eq:proba}
\end{equation}
where $Z_{ \vec {\lambda }}$ is the partition sum of this intensive
ensemble controlled by the Lagrange multipliers $ \vec {\lambda }{\equiv }%
\{\lambda _{k}^{(i)}\}$ and $\bar{W}(\vec{b})=Z_{{0}}P_{{0}}({ \vec{b}})$
is the partition sum of the statistical ensemble associated with fixed
values ${ \vec{b}}$ of all the observables. In the following we will call it
the extensive ensemble using the word ''extensive'' in the weak sense of
''observable''~\cite{balian}, meaning that observables are controlled on an
event-by-event basis in the extensive ensemble. The two partition sums are
related through the usual Laplace transform $Z_{{ \vec{\lambda} }}=\int d%
{ \vec{b}}\;\bar{W} ( { \vec{b}} ) \exp (-{ \vec {\lambda} 
\cdot \vec{b}})$. Eq.(%
\ref{eq:proba}) clearly demonstrates that the convexity anomalies of the
thermodynamical potential $\log \bar{W}$ can be traced back from $\log P_{%
{ \vec{\lambda} }} ( { \vec{b}} ) $. Indeed, the equation of states
of the extensive ensemble can be related to the derivative of $\log P_{%
{ \vec{\lambda} }} ( { \vec{b}} ) $ 
\begin{equation}
\bar{\lambda}_{k} ( { \vec{b}} ) \equiv \frac{\partial \log \bar{W}%
 ( { \vec{b}} ) }{\partial b_{k}}= \frac
{\partial \log P_{ \vec{\lambda} } ( \vec{b} )} 
{\partial b_{k}}
+\lambda _{k}.
\label{eq:alpha_bar}
\end{equation}
while the curvature matrix of $\log \bar{W}$ is the one of $\log P_{\vec 
{\lambda} } ( { \vec{b}} ) $

\begin{equation}
C_{kk^{\prime }} ( { \vec{b}} ) \equiv \frac{\partial \bar{\lambda}%
_{k}( { \vec{b}} ) }{\partial b_{k^{\prime }}}\equiv \frac{%
\partial ^{2}\log \bar{W} ( { \vec{b}} ) }{\partial b_{k}\partial
b_{k^{\prime }}}=\frac{\partial ^{2}\log P_{{ \vec{\lambda} }}
 ( \vec {b} ) }{\partial b_{k}\partial b_{k^{\prime }}}.
\end{equation}
It should be noticed that these relations are valid for every set of
Lagrange multipliers ${ \vec {\lambda} }$ because in a finite system the
probability distribution covers the entire accessible space of ${ \vec{b}}$
and so contains the whole thermodynamics of the extensive ensemble. The only
limitation can be a practical one for experiments or simulations: to
accumulate enough statistics at every location.

Let us consider the simplified case of a unique direction $b_{k}\equiv b$
and so a unique Lagrange multiplier $\lambda _{k}\equiv \lambda $. If $\bar{W%
}\left( b\right) $ has an abnormal curvature, then the EOS $\bar{\lambda}%
\left( b\right) =\partial _{b}\log \bar{W}\left(  b \right) $
presents a back-bending. For this extensive ensemble where $b$ is the
control parameter, coexistence can be defined as the region where $\bar{%
\lambda}$ is associated to three values of $b$ because of the anomalous
curvature. The definition of a first order phase transition as the occurrence
of a negative heat capacity~\cite{gauss,gross} is readily obtained if one
identifies $S=\log \bar{W}$ as the microcanonical entropy, $b$ as the energy
and $\lambda =\beta $ the inverse canonical temperature. The generalization
to the EOS of any thermodynamic potential involving an extensive variable $%
b$ \cite{prl99} is also obvious. For example, a bimodal grand-canonical
number of particle distribution is equivalent to a negative chemical
susceptibility in the canonical ensemble, a bimodal distribution of
magnetization to a negative magnetic susceptibility in the constant
magnetization ensemble, a bimodal volume distribution in the isobar ensemble
to a negative compressibility of the isochore ensemble, ...

For $\lambda $ in the region of anomalous curvature the associated
probability distribution presents two maxima and a minimum. In the intensive
ensemble eq.(\ref{eq:proba}) where $\lambda $ is controlled, 
coexistence is then signalled by the bimodality of the probability
distribution and the value of $\lambda $ for which the two maxima have equal
height is the first order transition point, analogous to the usual Maxwell
construction. Therefore, the definitions of a first order phase transition
through the occurrence of an abnormal curvature of the thermodynamical
potential of the extensive ensemble, or through the presence of a bimodal
event distribution in the associated intensive ensemble are strictly
equivalent.

Let us now consider the definition~\cite{grossmann} based on the zeroes of $%
Z_{{ \vec{\lambda} }}$, the partition sum of the intensive ensemble eq.(\ref
{eq:proba}), in the complex ${ \vec{\lambda} }$ plane. We shall 
now work out the relations between these zeroes and the concavity of $P_{%
{ \vec{\lambda} }}$. Let us continue for simplicity with a single observable 
$b_{k}\equiv b$ and the associated Lagrange multiplier $\lambda _{k}\equiv
\lambda $. Looking at the zeros of $Z_{\lambda }$ in the complex $\lambda $
plane, we first show that a bimodal distribution corresponds to a partition sum
fulfilling the Yang Lee theorem in the double saddle point approximation~\cite
{topology}. Then we will demonstrate the reciprocal: a distribution of zeroes
which fulfills the Yang Lee theorem is associated to a bimodal 
probability distribution for $b$ .

Using equation (\ref{eq:proba}) we see that the partition sum for a complex
parameter ${ \gamma }=\lambda +i\eta $ is nothing but the Laplace
transform of the probability distribution $P_{\lambda _{0}}\left( b\right) $
for any arbitrary parameter $\lambda _{0}$ \cite{lee} 
\begin{equation}
Z_{{ \gamma }}=\int dbZ_{\lambda _{0}}\;P_{\lambda _{0}}\left(
b\right) e^{-\left( { \gamma }-\lambda _{0}\right) b}\equiv \int
db\;p_{\lambda _{0}}\left( b\right) e^{-\bar{{ \gamma }}b}
\label{laplace}
\end{equation}
where $\bar{{ \gamma }}={ \gamma }-\lambda _{0}$ and we have
defined $p_{\lambda }\left( b\right) =Z_{\lambda }\;P_{\lambda }\left(
b\right) $. We can also write the partition sum at a complex point 
${\gamma }=\lambda +i\eta $ as a Fourier transform of the probability
distribution $p_{\lambda }\left( b\right) $%
\begin{equation}
Z_{{ \gamma }}=\int db\;p_{\lambda }\left( b\right) e^{-i\eta b}
\label{fourier}\end{equation}
If $P_{\lambda }\left( b\right) $ is monomodal we can use a saddle point
approximation around the maximum $\bar{b}_{\lambda }$ giving 
$Z_{\gamma}=e^{\phi _{{ \gamma }}\left( \bar{b}_{\lambda }\right) }$ ,
with 
\begin{equation}
\phi _{{ \gamma }}\left( b\right) =\log p_{\lambda }\left( b\right)
-i\eta b+ \frac{1}{2}\eta ^{2}\sigma ^{2}\left( b\right) 
+\frac{1}{2}\log \left( 2\pi \sigma
^{2}(b)\right)  \label{saddle}
\end{equation}
where $\sigma ^{-2}=\partial _{b}^{2}\log p_{\lambda _{0}}\left( b\right) $.
However, if $\bar{W}_{\lambda _{0}}(b)$ has a curvature anomaly it exists a
range of $\lambda $ for which the equation $\partial _{b}\log \bar{W}%
_{\lambda _{0}}=\lambda -\lambda _{0}$ has three solutions, one minimum and
two maxima $b_{\lambda }^{(1)}$and $b_{\lambda }^{(2)}$ . Let us first split
the probability distribution into two normal components 
\begin{equation}
p_{\lambda }(b)=p_{\lambda }^{(1)}(b)+p_{\lambda }^{(2)}(b)
\end{equation}
$p_{\lambda }^{(i)}(b)$ being peaked at $b_{\lambda }^{(i)}.$ The partition
sum reads 
\begin{equation}
Z_{{ \gamma }}=\,e^{\phi _{{ \gamma }}^{(1)}}+\,e^{\phi _{%
{ \gamma }}^{(2)}}=2e^{\phi _{{ \gamma }}^{+}}\cosh (\phi _{%
{ \gamma }}^{-})
\end{equation}
where $2\phi _{{ \gamma }}^{+}=\phi _{{ \gamma }}^{(1)}+\phi _{%
{ \gamma }}^{(2)}$ and $2\phi _{{ \gamma }}^{-}=\phi _{ 
\gamma }^{(1)}-\phi _{{ \gamma }}^{(2)}$. We can now use a
double saddle point approximation around the two maxima which will be valid
close to thermodynamical limit~\cite{lee} and we get $\phi _{{ \gamma }%
}^{\left( i\right) }=\phi _{{ \gamma }}(b_{\lambda }^{(i)})\ $%
according to eq. (\ref{saddle}). The zeros of $Z_{{ \gamma }}$ then
correspond to 
\begin{equation}
2\phi _{{ \gamma }}^{-}=i(2n+1)\pi
\end{equation}
The imaginary part is given by $\eta =(2n+1)\pi /\Delta b$ where $\Delta
b=b_{\lambda }^{(2)}-b_{\lambda }^{(1)}$ is the jump in $b$ between the two
phases. For the real part we should solve the equation $\Re (\phi _{{ \bf %
\gamma }}^{-})=0.$ In particular, close to the real axis this equation
defines a $\lambda $ which can be taken as $\lambda _{0}$. If the bimodal
structure persists when the number of particles goes to infinity, the loci
of zeros corresponds to a line perpendicular to the real axis with a uniform
distribution as expected for a first order phase transition~\cite{yanglee}.

Let us now work out the necessary condition and show that a uniform
distribution of zeroes perpendicular to the real axis with a density
linearly increasing with the number of particles $N$ implies a bimodal 
probability distribution.
Let us denote zeros as 
\begin{equation}
{ \gamma }_{n}=\lambda _{0}+i(2n+1)\frac{\pi} {N \delta}   \label{zeroes}
\end{equation}
such that the interval $2\pi /\delta $ contains $N$ uniformly distributed
zeroes in agreement with the unit circle theorem~\cite{yanglee}. Since all
the zeros of $Z_{{ \gamma }}$ are periodically distributed one can
define an analytic function $f({ \gamma })$ such that 
\begin{equation}
Z_{{ \gamma }}=2\cosh \left( \left( { \gamma }-\lambda
_{0}\right) \frac{N\delta} {2}\right) f({ \gamma })  \label{fourier2}
\end{equation}
Indeed since $Z_{{ \gamma }}$ is analytic, $f({ \gamma })$ could in principle
diverge only on the zeroes of $Z_{{ \gamma }}$; in these points however
eq.(\ref{fourier2}) shows that $f$ is proportional to 
$\partial_{\gamma} Z_{\gamma}$, $i.e.$ it is analytic.
On the line of zeroes, 
relation (\ref{fourier2}) reduces to $Z_{\lambda _{0}+i\eta
}=2\cos \left( \eta N\delta /2\right) f(\lambda _{0}+i\eta ).$ Using the
fact that the partition sum along the line of zeroes is the Fourier
transform of the reduced probability distribution at the transition point $%
\lambda _{0}$ (eq.(\ref{fourier}),
we can use the inverse Fourier transform to get the distribution \thinspace $%
p_{\lambda _{0}}\left( b\right) $ at the transition point  
\begin{equation}
p_{\lambda _{0}}\left( b\right) = \frac{1}{2 \pi}\int d\eta Z_{\lambda
_{0}+i\eta }\;e^{i\eta b}  \label{p}
\end{equation}
Equation (\ref{p}) can then be rewritten as  
\begin{equation}
p_{\lambda _{0}}\left( b\right) =g_{\lambda _{0}}
\left( b+\frac{N\delta} {2}\right)
+g_{\lambda _{0}}\left( b-\frac{N\delta} {2}\right)   \label{bimod}
\end{equation}
where 
\[
g_{\lambda }(b)= \frac{1}{2\pi}\int d\eta f( \gamma)\;e^{i\eta b}
\]
is a distribution. Indeed if we compute $p_{\lambda}(b)$ a little above 
$\lambda _{0}$, $\lambda =\lambda _{0}+\varepsilon $ we get 
\begin{equation}
p_{\lambda _{0}+\varepsilon }\left( b\right) =e^{\varepsilon N\delta
/2}g_{\lambda _{0}+\varepsilon }\left( b+\frac{N\delta} {2}\right) 
+e^{-\varepsilon
N\delta /2}g_{\lambda _{0}+\varepsilon }\left( b-\frac{N\delta} {2}\right) 
\label{coexistence}
\end{equation}
For large $N$ only the first term survives

\begin{equation}
p_{\lambda _{0}+\varepsilon }\left( b\right) \simeq e^{\varepsilon N\delta
/2}g_{\lambda _{0}+\varepsilon }\left( b+\frac{N\delta} {2}\right) 
\equiv p_{\lambda
_{0}+\varepsilon }^{\left( 1\right) }\left( b\right)
\end{equation}
which is the first term in the distribution at the transition point eq.(\ref
{p}). If we conversely compute $p_{\lambda}( b) $ 
a little below the transition point we get

\begin{equation}
p_{\lambda _{0}-\varepsilon }\left( b\right) \simeq e^{\varepsilon N\delta
/2}g_{\lambda _{0}-\varepsilon }\left( b-\frac{N\delta} {2}\right) 
\equiv p_{\lambda
_{0}-\varepsilon }^{\left( 2\right) }\left( b\right)
\end{equation}
which is the second term in the transition point distribution eq.(\ref{p}).
At the transition point  $p_{\lambda _{0}}\left( b\right) $
is the sum of two shifted (identical) distributions, $p_{\lambda _{0}}^{\left(
1\right) }$ and $p_{\lambda _{0}}^{\left( 2\right) }$.
If $g_\lambda$ is a normal ($i.e.$ monomodal) 
distribution the central limit theorem guarantees
that for a large number of particles its width will scale as $\sqrt{N}$,
$i.e.$ will grow slower than the distance between the two peaks that scales as
$N$. This implies that (for not too small $N$) 
$p_{\lambda _{0}}\left( b\right)$ 
is bimodal. This stays true if $g_\lambda$ itself is bimodal
(or multimodal), with the only difference that the total distribution  
$p_{\lambda _{0}}\left( b\right)$ will then represent the coexistence between
more than two phases. 
Let consider the simplest case of a bimodal structure for 
$p_{\lambda _{0}}\left( b\right) $, close to the
thermodynamical limit; a little before the transition point only the lowest
peak of the $b$ distribution is present, while a little above only the
second one remains. At the transition both are present and the most probable 
$b$ jumps from $b_{<}=\bar{b}+N\delta /2$ to $b_{>}=\bar{b}-N\delta /2$
where $\bar{b}$ is the maximum of $g_{\lambda _{0}}\left( b\right) .$ $%
\delta $ represents the discontinuity of the variable $b$ per particle. If $%
b $ is the energy $\delta $ is the latent heat per particle.

\smallskip

The important difference between finite systems and the thermodynamic limit
is that in the latter case the probability distribution is bimodal only at
the transition point $\lambda _{0}$. On the other hand in finite systems, in
an interval $\Delta \lambda $ of non-zero measure around the transition
point $\lambda _{0}$ the two phases coexist, $i.e.$ the distribution is
bimodal, each peak being associated with a phase having a finite probability
of occurrence. Eq.(\ref{coexistence}) shows that this interval is the larger
the smaller is $\delta $, $i.e.$ the lower is the density of zeroes. This
extension of the phase transition point to a region in which the intensive
ensemble presents a bimodal distribution makes a direct measurement of phase
coexistence possible \cite{lopezo}, contrary to the common belief that phase
transitions would only be loosely defined out of the thermodynamic limit.  
Moreover, the fact that the distribution is non zero between the two maxima 
$b_{<}$
and $b_{>}$ in the case of small systems (see eq.(\ref{coexistence}))
implies that  
microstates can be accessed that do not exist at the
thermodynamic limit in the intensive ensemble. These states are specific of
the coexistence region and can lead to spectacular phenomena as negative
heat capacity~\cite{gauss} and negative compressibility~\cite{prl99} in the
extensive ensemble.

\smallskip

\smallskip This demonstration can be extended to account for more important 
finite
size effects~\cite{lee} when zeroes ${ \gamma }_{n}$ are still periodic 
but not yet
uniformly distributed on a straight line~\cite{lee,grossmann}. Indeed, it
is important to remark that the distribution of zeroes is periodic in the
imaginary direction before the thermodynamic limit. If for example we
consider the grancanonical ensemble, the Lagrange parameter $\lambda \equiv
-\alpha \equiv -\beta \mu $ represents the logarithm of the fugacity $%
z=e^{\alpha }$ and the observable is the number of particles $b\equiv N$. The
partition sum $Z_{\alpha }\propto e^{\alpha N}$ in the complex $\alpha $
plane is periodic for any arbitrary value of $N$. In the more
general case, this will not be true for  very small systems since the
extensive variables $b_{k}$ are not in general proportional to the number of
constituents of the system. However, if the forces are short-ranged and $N$
is sufficiently high, $b_{k}\propto N$, which guarantees the periodicity of 
$Z_{\lambda}$ in the complex $\lambda$ plane. This
periodicity is a constraint on the distribution of zeroes.

To take into account finite size distortions, we can introduce
a transformation 
\begin{equation}
m({ \gamma })={ \gamma }-\Delta m({ \gamma })
\end{equation}
which is the identity on the real axis and which maps the zeroes 
${\gamma }_{n}$ of the
partition sum onto a uniform density perpendicular to the real axis 
\begin{equation}
m({ \gamma }_{n})=\lambda _{0}+i(2n+1)\frac{\pi} {N\delta} 
\end{equation}
As before we can introduce an analytic function 
$\bar{f}({ \gamma })$ such that 
\begin{equation}
Z_{{ \gamma }}=2\cosh \left( \left( m\left( { \gamma }\right)
-\lambda _{0}\right) \frac{N\delta} {2}\right) \bar{f}({ \gamma })
\end{equation}
 Using this expression in the inverse Fourier transform eq.(\ref{p}) the
probability distribution at the transition point reads 
\begin{equation}
p_{\lambda _{0}}(b)=\bar{g}_{\lambda _{0}}^{+}(b+\frac{N\delta} {2})+\bar{g}%
_{\lambda _{0}}^{-}(b-\frac{N\delta} {2})  \label{p1}
\end{equation}
where 
\begin{equation}
\bar{g}_{\lambda _{0}}^{\pm }(b)={    N}_{0}\int d\eta \bar{f}(\lambda
_{0}+i\eta )e^{\mp \Delta m(\lambda _{0}+i\eta )N\delta /2}e^{i\eta b}
\end{equation}
If a phase transition is present at the thermodynamic limit, according to
the Yang-Lee theorem $\Delta m$ goes to zero faster than $N$ such that 
the two
$\bar{g}_{\lambda _{0}}^{\pm }$ converge to the same distribution 
$g_{\lambda _{0}}$,
and $p_{\lambda _{0}}(b)$ tends to eq.(\ref{bimod}). Following the 
discussion of eq.(\ref{p1}), 
this means that a deformed distribution of zeroes that converges
to a straight line of equally spaced zeroes perpendicular to the real axis
gives rise to a bimodal probability distribution function.




In finite systems, if $\Delta m$ is not zero,  the sum (\ref{p1}) may or may
not present a concavity anomaly depending on the actual properties of $%
\Delta m$. This is true irrespectively of the fact that the straight
distribution of zeroes eq.(\ref{zeroes}) is asymptotically reached ($i.e.$ a
first order phase transition exists in the bulk) or not. In this case, the
physical phenomenon could be classified differently with the zeroes of the
partition sum~\cite{grossmann} or with the concavity criteria~\cite
{gross,prl99}. Indeed, we have demonstrated their equivalence only for large
systems.

In conclusion, our topological definition of phase transitions through the
bimodality of the probability distribution function, 
generalizes the definition based on the
microcanonical entropy \cite{gauss,gross} and is equivalent to the definition
based on convexity anomalies of generic thermodynamic potentials \cite
{prl99} for any number of particles. We have demonstrated that a bimodality
is equivalent to the Yang Lee definition of phase transitions close to and
at the thermodynamical limit. A uniform distribution of zeroes perpendicular
to the real axis of a Lagrange multiplier $\lambda $ at the position $%
\lambda _{0}$ implies that the associated distribution of events is bimodal
for this $\lambda _{0}$ , and conversely a bimodal distribution at a given $%
\lambda _{0}$ generates a uniform distribution of zeroes with a real part $%
\lambda _{0}$ and a regularly spaced imaginary part. This means that the
occurrence of bimodalities can be considered as a valid generalization of
the concept of phase transition to systems of any size. This topological
definition is extremely powerful since it gives access to the order
parameters defined as the observables for which the distribution
is bimodal. We have
demonstrated that this definition agrees with the definition based on the
zeroes of $Z$~\cite{grossmann} when the bimodal distribution can be
approximated by two gaussians, i.e. in the double saddle point
approximation. This means that the different definitions are coherent for $N$
big enough that fluctuations around the two phases are at the gaussian
level, but they can differ for smaller systems. The compatibility of the
different definitions at the thermodynamic limit means that the great effort
done by the different research groups to recognize and classify phase
transitions in finite systems~\cite{grossmann,gross,noi} is not only rigorous
but also consistent with macroscopic thermodynamics.

\end{document}